\documentclass[preprint,aps,superscriptaddress,nofootinbib,tightenlines]{revtex4}
\usepackage{epsfig}
\usepackage{bm}

\usepackage{epsfig}
\newcommand{\beq}{\begin{equation}}
\newcommand{\eeq}{\end{equation}}
\newcommand{\bea}{\begin{eqnarray}}
\newcommand{\eea}{\end{eqnarray}}

\def\OMIT#1{{}}

\newcommand{\lsim}{\raisebox{-0.7ex}{$\stackrel{\textstyle <}{\sim}$ }}

\begin{document}
\begin{figure}[!t]
\vskip -1.5cm
\leftline{
{\epsfxsize=1.5in \epsfbox{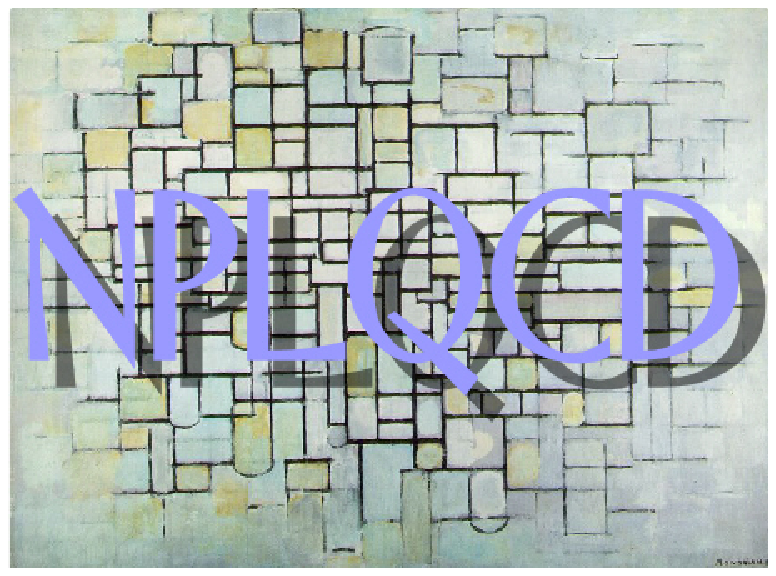}}}
\vskip 1.2cm
\end{figure}

\preprint{\vbox{
\hbox{UNH-07-01}
\hbox{UCRL-JRNL-231719}
\hbox{UMD-40762-391}
\hbox{JLAB-THY-07-653}
\hbox{NT@UW-07-08}
}}

\vskip .5cm

\title{Precise Determination of the $I=2$ $\pi\pi$ Scattering Length\\  from Mixed-Action Lattice QCD}

\vskip .5cm
\author{Silas R.~Beane}
\affiliation{Department of Physics, University of New Hampshire,
Durham, NH 03824-3568.}
\author{Thomas C.~Luu}
\affiliation{N Division, Lawrence Livermore National Laboratory, Livermore, CA 94551.}
\author{Kostas Orginos}
\affiliation{Department of Physics, College of William and Mary, Williamsburg,
  VA 23187-8795.}
\affiliation{Jefferson Laboratory, 12000 Jefferson Avenue, 
Newport News, VA 23606.}
\author{Assumpta Parre\~no}
\affiliation{Departament d'Estructura i Constituents de la Mat\`{e}ria and
Institut de Ci\`encies del Cosmos, 
Universitat de Barcelona,  E--08028 Barcelona, Spain.}
\author{Martin J.~Savage}
\affiliation{Department of Physics, University of Washington, 
Seattle, WA 98195-1560.}
\author{Aaron Torok}
\affiliation{Department of Physics, University of New Hampshire,
Durham, NH 03824-3568.}
\author{Andr\'e Walker-Loud}
\affiliation{Department of Physics, University of Maryland, College Park, MD 20742-4111.}
%\qquad}
\collaboration{ NPLQCD Collaboration }
\noaffiliation
\vphantom{}
\vskip 0.8cm
\begin{abstract}
\noindent 
The $I=2$ $\pi\pi$ scattering length is calculated in fully-dynamical
lattice QCD with domain-wall valence quarks on the asqtad-improved
coarse MILC configurations (with fourth-rooted staggered sea quarks)
at four light-quark masses.  Two- and three-flavor mixed-action chiral
perturbation theory at next-to-leading order is used to perform the
chiral and continuum extrapolations.  At the physical charged pion
mass, we find $m_\pi a_{\pi\pi}^{I=2} = -0.04330 \pm 0.00042$, where
the error bar combines the statistical and systematic uncertainties in
quadrature.
\end{abstract}
\maketitle

%\vfill\eject

%\tableofcontents

\vfill\eject

%%%%%%%%%%%%%%%%%%%%%%%%%%%%%%%%%%%%%%%%%
\section{Introduction}
\label{sec:Intro}

\noindent Pion-pion ($\pi\pi$) scattering at low energies is the
simplest and best-understood hadron-hadron scattering process.  Its
simplicity and tractability follow from the fact that the pions are
identified as the pseudo-Goldstone bosons associated with the
spontaneous breaking of the approximate chiral symmetry of Quantum
Chromodynamics (QCD).  For this reason, the low-momentum interactions
of pions are strongly constrained by the approximate chiral
symmetries, more so than other hadrons.  The scattering lengths for
$\pi\pi$ scattering in the s-wave are uniquely predicted at leading
order (LO) in chiral perturbation theory
($\chi$-PT)~\cite{Weinberg:1966kf}:  
\begin{eqnarray}
m_\pi a_{\pi\pi}^{I=0} \ = \ 0.1588 \ \ ; \ \ m_\pi a_{\pi\pi}^{I=2} \ = \
-0.04537 
\ \ \ ,
\label{eq:CA}
\end{eqnarray}
at the charged pion mass.
Subleading orders in the chiral expansion of the $\pi\pi$ amplitude
give rise to perturbatively-small deviations from the tree level, and
contain both calculable non-analytic contributions and analytic terms
with new coefficients that are not determined by chiral symmetry
alone~\cite{Gasser:1983yg,Bijnens:1995yn,Bijnens:1997vq}.  In order to
have predictive power at subleading orders, these coefficients must be
obtained from experiment or computed with lattice QCD.

Recent experimental efforts have been made to compute the s-wave
$\pi\pi$ scattering lengths, $a_{\pi\pi}^{I=0}$ ($I=0$) and
$a_{\pi\pi}^{I=2}$ ($I=2$): E865~\cite{Pislak:2001bf,Pislak:2003sv}
($K_{e4}$ decays), CERN DIRAC~\cite{Adeva:2005pg} (pionium lifetime)
and CERN NA48/2~\cite{Batley:2005ax}
($K^\pm\rightarrow\pi^\pm\pi^0\pi^0$).  Unfortunately, these
experiments do not provide stringent constraints on
$a_{\pi\pi}^{I=2}$.  However, a theoretical determination of s-wave
$\pi\pi$ scattering lengths which makes use of experimental data has
reached a remarkable level of
precision~\cite{Colangelo:2001df,Leutwyler:2006qq}:
\begin{eqnarray}
m_\pi a_{\pi\pi}^{I=0} \ = \ 0.220\pm 0.005 \ \ ; \ \ m_\pi a_{\pi\pi}^{I=2} \ = \ -0.0444\pm 0.0010
\ \ \ .
\label{eq:roy}
\end{eqnarray}
These values result from the Roy
equations~\cite{Roy:1971tc,Basdevant:1973ru,Ananthanarayan:2000ht},
which use dispersion theory to relate scattering data at high energies
to the scattering amplitude near threshold. In a striking recent
result, this technology has allowed a model-independent determination
of the mass and width of the resonance with vacuum quantum numbers
(the $\sigma$ meson) that appears in the $\pi\pi$ scattering
amplitude~\cite{Caprini:2005zr}. Several low-energy constants of
one-loop $\chi$-PT are critical inputs to the Roy equation analysis.
One can take the values of these low-energy constants computed with
lattice QCD by the MILC
collaboration~\cite{Aubin:2004fs,Bernard:2006wx} as inputs to the Roy
equations, and obtain results for the scattering lengths consistent
with the analysis of Ref.~\cite{Colangelo:2001df}.

A direct lattice QCD determination of threshold $\pi\pi$ scattering is
problematic in two respects. First, the occurrence of disconnected
diagrams in the $I=0$ s-wave channel renders a determination of that
amplitude very costly in terms of computer time, given the current
state of lattice algorithms, and is thus beyond our current
capabilities.  As a result, lattice QCD efforts have focused on the
$I=2$ channel.  The second difficulty is due to the fact that lattice
QCD calculations are performed on a Euclidean lattice.  The
Maiani-Testa theorem demonstrates that S-matrix elements cannot be
determined from lattice calculations of $n$-point Green's functions at
infinite volume, except at kinematic thresholds~\cite{Maiani:1990ca}.
This difficulty was overcome by L\"uscher, who showed that by
computing the energy levels of two-particle states in the
finite-volume lattice, the $2\rightarrow 2$ scattering amplitude can
be
recovered~\cite{Huang:1957im,Hamber:1983vu,Luscher:1986pf,Luscher:1990ux,Beane:2003da}.
The energy levels of the two interacting particles are found to
deviate from those of two non-interacting particles by an amount that
depends on the scattering amplitude and varies inversely with the
lattice spatial volume.

The first lattice calculations of $\pi\pi$ scattering were
performed in quenched
QCD~\cite{Sharpe:1992pp,Gupta:1993rn,Kuramashi:1993ka,Kuramashi:1993yu,Fukugita:1994na,Gattringer:2004wr,Fukugita:1994ve,Fiebig:1999hs,Aoki:1999pt,Liu:2001zp,Liu:2001ss,Aoki:2001hc,Aoki:2002in,Aoki:2002sg,Aoki:2002ny,Juge:2003mr,Ishizuka:2003nb,Aoki:2005uf,Aoki:2004wq,Li:2007ey},
and 
the first full-QCD calculation of $\pi\pi$ scattering (the scattering length
and phase-shift) was carried
through by the CP-PACS collaboration, who exploited the finite-volume
strategy to study $I=2$, s-wave scattering with two flavors ($n_f=2$)
of
improved Wilson fermions~\cite{Yamazaki:2004qb}, with pion masses in
the range $m_\pi\simeq 0.5-1.1~{\rm GeV}$.  
The first fully-dynamical
calculation of the $I=2$ $\pi\pi$ scattering length with three flavors ($n_f=2+1$)
of light quarks was performed by some of the present authors using domain-wall
valence quarks on asqtad-improved staggered sea quarks at four pion
masses in the range $m_\pi\simeq 0.3-0.5~{\rm GeV}$ at a single
lattice spacing, $b\sim 0.125~{\rm fm}$ ~\cite{Beane:2005rj}. That
work quoted a value of the scattering length extrapolated to the physical point
of
\begin{eqnarray}
m_\pi a_{\pi\pi}^{I=2} & = & -0.0426\pm 0.0006 \pm 0.0003\pm 0.0018
\ \ \ ,
\label{eq:nplqcd1}
\end{eqnarray}
where the first uncertainty is statistical, the second is a
systematic due to fitting and the third
uncertainty is due to truncation of the chiral expansion.

In this paper we update our fully-dynamical mixed-action
calculation of the $I=2$ $\pi\pi$ scattering
length. Two recent developments motivate an update: 
i) we have vastly increased statistics at the three light-quark
masses studied in the original publication; 
ii) $\pi\pi$ scattering
has been computed with Mixed-Action $\chi$-PT (MA$\chi$-PT) at next-to-leading
order (NLO)~\cite{Chen:2005ab,Chen:2006wf} both for two and three flavors of light
quarks. Our updated result is:
\begin{eqnarray}
m_\pi a_{\pi\pi}^{I=2} & = &  -0.04330 \pm 0.00042
\ \ \ ,
\label{eq:nplqcd2}
\end{eqnarray}
where the statistical and systematic uncertainties have been combined in quadrature.
This result is consistent with all previous determinations within uncertainties.

This paper is organized as follows.  In Section~\ref{sec:Method}
details of our mixed-action lattice QCD calculation are presented. We
refer the reader interested in a more comprehensive treatment and
discussion to our earlier papers.  Discussion of the relevant
correlation functions and an outline of the methodology and fitting
procedures can also be found in this section.  
The results of the lattice calculation
and the analysis with two- and three-flavor MA$\chi$-PT are presented
in Section~\ref{sec:Extrapolate}. In Section~\ref{sec:Systerrors}, the
various sources of systematic uncertainty are identified and quantified.  In
Section~\ref{sec:Conclude} we conclude.

%%%%%%%%%%%%%%%%%%%%%%%%%%%%%%%%%%%%
\section{Methodology and Details of the Lattice Calculation}
\label{sec:Method}

\noindent 
The computation in this paper uses the mixed-action lattice QCD scheme
developed by LHPC~\cite{Renner:2004ck,Edwards:2005kw}.  Domain-wall
fermion propagators were generated from a smeared source on $n_f=2+1$
asqtad-improved~\cite{Orginos:1999cr,Orginos:1998ue} coarse
configurations generated with rooted staggered sea
quarks~\cite{Bernard:2001av}. Hypercubic-smeared
(HYP-smeared)~\cite{Hasenfratz:2001hp,DeGrand:2002vu,DeGrand:2003in,Durr:2004as}
gauge links were used in the domain-wall fermion action to improve
chiral symmetry (further details about the mixed-action scheme can be
found in Refs.~\cite{Beane:2006gf,Beane:2006gj}).  The mixed-action
calculations we have performed involved computing the valence-quark
propagators using the domain-wall formulation of lattice fermions, on
each gauge-field configuration of an ensemble of the coarse MILC
lattices that are generated using the staggered formulation of lattice
fermions~\cite{Kaplan:1992bt,Shamir:1992im,Shamir:1993zy,Shamir:1998ww,Furman:1994ky}
and taking the fourth root of the fermion determinant,
i.e. domain-wall valence quarks on a rooted-staggered sea.  In the
continuum limit the $n_f=2$ staggered action has an $SU(8)_L\otimes
SU(8)_R\otimes U(1)_V$ chiral symmetry due to the four-fold taste
degeneracy of each flavor, and each pion has 15 degenerate additional
partners.  At finite lattice spacing this symmetry is broken and the
taste multiplets are no longer degenerate, but have splittings that are 
${\cal O}(\alpha^2 b^2)$. While there is no proof, there are arguments to
suggest that taking the fourth root of the fermion determinant
recovers the contribution from a single Dirac fermion~\footnote{For a
nice introduction to staggered fermions and the fourth-root trick, see
Ref.~\cite{degrandANDdetar}.  For the most recent discussions
regarding the continuum limit of staggered fermions with the
fourth-root trick, see Ref.~\cite{Durr:2004ta,Creutz:2006ys,Bernard:2006vv,Durr:2006ze,Hasenfratz:2006nw,Bernard:2006ee,Shamir:2006nj,Sharpe:2006re}.}. The results of
this paper assume that the fourth-root trick recovers the correct
continuum limit of QCD.

When determining the mass of the valence quarks there is an ambiguity
due to the non-degeneracy of the 16 staggered bosons associated with
each pion. One could choose to match to the taste-singlet meson or to
any of the mesons that become degenerate in the continuum limit.
Given that the effective field theory exists to describe such
calculations at finite lattice spacing, the effects of matching can be
described, and removed, by effective field theory calculations
appropriate to the choice of matching.  The quantity $b^2\Delta_I$ is
the mass-difference between a valence meson and the staggered
taste-singlet meson when the valence pion is tuned to be exactly
degenerate with the lightest staggered pion.  On the coarse MILC
lattices with $b\sim 0.125~{\rm fm}$ (and $L\sim 2.5~{\rm fm}$) it is
numerically determined (in lattice units) that $b^2\Delta_I\ =\
0.0769(22)$~\cite{Aubin:2004fs}.

A summary of the lattice parameters and resources used in this work is
given in Table~\ref{tab:MILCcnfs}.  In order to generate large
statistics on the existing MILC configurations, multiple propagators
from sources displaced both temporally and spatially on the lattice
were computed.  The correlators were blocked so that one average
correlator per configuration was used in the subsequent Jackknife
statistical analysis (that will be described later).
\begin{table}[!tb]
 \caption{The parameters of the MILC gauge configurations and
   domain-wall propagators used in this work. The subscript $l$
   denotes light quark (up and down), and  $s$ denotes the strange
   quark. The superscript $dwf$ denotes the bare-quark mass for the
   domain-wall fermion propagator calculation. The last column is the 
   number of configurations times the number of sources per
   configuration.}
\label{tab:MILCcnfs}
\begin{ruledtabular}
\begin{tabular}{ccccccc}
 Ensemble        
&  $b m_l$ &  $b m_s$ & $b m^{dwf}_l$ & $ b m^{dwf}_s $ & $10^3 \times b
m_{res}$~\protect\footnote{Computed by the LHP collaboration.} & \# of propagators   \\
\hline 
2064f21b676m007m050 &  0.007 & 0.050 & 0.0081 & 0.081  & $1.604\pm 0.038$ & 468\ $\times$\ 16 \\
2064f21b676m010m050 &  0.010 & 0.050 & 0.0138 & 0.081  & $1.552\pm 0.027$ & 658\ $\times$\ 20 \\
2064f21b679m020m050 &  0.020 & 0.050 & 0.0313 & 0.081  & $1.239\pm 0.028$ & 486\ $\times$\ 24 \\
2064f21b681m030m050 &  0.030 & 0.050 & 0.0478 & 0.081  & $0.982\pm 0.030$ & 564\ $\times$\ 8 \\
\end{tabular}
\end{ruledtabular}
\end{table}

The $\pi$ correlation function, $C_\pi (t)$, and the $\pi\pi$
correlation function $C_{\pi\pi} (p,t)$ were computed, where the
number of time slices between the hadronic sink and the hadronic
source is denoted by $t$, and $p$ denotes the magnitude of the (equal
and opposite) momentum of each pion.  The single-$\pi^+$ correlation
function is
\begin{eqnarray}
C_{\pi^+}(t) & = & \sum_{\bf x}
\langle \pi^-(t,{\bf x})\ \pi^+(0, {\bf 0})
\rangle
\ \ \ ,
\label{pi_correlator} 
\end{eqnarray}
where the summation over ${\bf x}$ corresponds to summing over all the
spatial lattice sites, thereby projecting onto the momentum ${\bf
p}={\bf 0}$ state.  A $\pi^+\pi^+$ correlation function that projects
onto the s-wave state in the continuum limit is
\begin{eqnarray}
C_{\pi^+\pi^+}(p, t) & = & 
\sum_{|{\bf p}|=p}\ 
\sum_{\bf x , y}
e^{i{\bf p}\cdot({\bf x}-{\bf y})} 
\langle \pi^-(t,{\bf x})\ \pi^-(t, {\bf y})\ \pi^+(0, {\bf 0})\ \pi^+(0, {\bf 0})
\rangle
\ \ \ , 
\label{pipi_correlator} 
\end{eqnarray}
where, in eqs.~(\ref{pi_correlator}) and (\ref{pipi_correlator}),
$\pi^+(t,{\bf x}) = \bar u(t, {\bf x}) \gamma_5 d(t, {\bf x})$ is an
interpolating field (Gaussian-smeared) for the $\pi^+$.
In the relatively large lattice volumes that we are using, the energy
difference between the interacting and non-interacting two-meson states
is a small fraction of the total energy, which is dominated by the
masses of the mesons.  In order to extract this energy difference we
formed the ratio of correlation functions, $G_{\pi^+ \pi^+}(p, t)$, where
\begin{eqnarray}
G_{\pi^+ \pi^+}(p, t) & \equiv &
\frac{C_{\pi^+\pi^+}(p, t)}{C_{\pi^+}(t) C_{\pi^+}(t)} 
\ \rightarrow \ \sum_{n=0}^\infty\ {\cal A}_n\ e^{-\Delta E_n\ t} 
\  \ ,
\label{ratio_correlator} 
\end{eqnarray}
and the arrow denotes the large-time behavior of $G_{\pi^+ \pi^+}$ in
the absence of boundaries on the lattice and becomes an equality in
the limit of an infinite number of gauge configurations.  In $G_{\pi^+
\pi^+}$, some of the fluctuations that contribute to both the one- and
two-meson correlation functions cancel, thereby improving the quality
of the extraction of the energy difference beyond what we are able to
achieve from an analysis of the individual correlation functions.

The energy eigenvalue $E_n$ and its deviation from the sum of the rest
masses of the particle, $\Delta E_n$, are related to the
center-of-mass momentum $p_n$ by
\begin{eqnarray}
\Delta E_n \ & \equiv & E_n\ -\  2 m_\pi \ = \ 2\sqrt{\ p_n^2\ +\ m_\pi^2\ } 
\ -\ 2m_\pi \ .
\label{eq:energieshift}
\end{eqnarray}
In the absence of interactions between the particles,
$|p\cot\delta|=\infty$, and the energy levels occur at momenta ${\bf
p} =2\pi{\bf j}/L$, corresponding to single-particle modes in a cubic
volume.  In the interacting theory, once the energy shift has been
computed, the real part of the inverse scattering amplitude is
determined via the L\"uscher
formula~\cite{Huang:1957im,Hamber:1983vu,Luscher:1986pf,Luscher:1990ux}.
To obtain $p\cot\delta(p)$, where $\delta(p)$ is the phase shift, the
magnitude of the center-of-mass momentum, $p$, is extracted from the
energy shift, given in eq.~(\ref{eq:energieshift}), and inserted
into~\cite{Huang:1957im,Hamber:1983vu,Luscher:1986pf,Luscher:1990ux,Beane:2003da}:
\begin{eqnarray}
p\cot\delta(p) \ =\ {1\over \pi L}\ {\bf
  S}\left(\,\frac{p L}{2\pi}\,\right)
\ \ ,
\label{eq:energies}
\end{eqnarray}
which is valid below the inelastic threshold. The regulated three-dimensional sum is~\cite{Beane:2003da}
\begin{eqnarray}
{\bf S}\left(\,{\eta}\, \right)\ \equiv \ \sum_{\bf j}^{ |{\bf j}|<\Lambda}
{1\over |{\bf j}|^2-{\eta}^2}\ -\  {4 \pi \Lambda}
\ \ \  ,
\label{eq:Sdefined}
\end{eqnarray}
where the summation is over all triplets of integers ${\bf j}$ such that $|{\bf j}| < \Lambda$ and the
limit $\Lambda\rightarrow\infty$ is implicit.
The approximate formula~\cite{Huang:1957im,Hamber:1983vu,Luscher:1986pf,Luscher:1990ux}
that can be used for $L\gg a$ is 
\begin{eqnarray}
\Delta E_0 &  = &  -\frac{4\pi a}{m_\pi  L^3}
\left[\ 1\ +\  c_1 \frac{a}{L}\ +\  c_2 \left( \frac{a}{L} \right)^2 \ \right ]
\ +\ {\cal O}\left({1\over L^6}\right)
\ \ ,
\label{luscher_a}
\end{eqnarray}
which relates the ground-state energy shift to the phase shift, with 
\begin{eqnarray}
c_1 & = & {1\over \pi}
\sum_{{\bf j}\ne {\bf 0}}^{ |{\bf j}|<\Lambda}
{1\over |{\bf j}|^2}\ -\   4 \Lambda \
\ =\ -2.837297
\ \ \ ,\ \ \
c_2\ =\ c_1^2 \ -\ {1\over \pi^2} \sum_{{\bf j}\ne {\bf 0}}
{1\over |{\bf j}|^4}
\ =\ 6.375183
\ ,
\end{eqnarray}
and $a$ is the scattering length, defined by
\begin{eqnarray}
a & = & \lim_{p\rightarrow 0}\frac{\tan\delta(p)}{p} 
\ \ \ .
\label{eq:scatt}
\end{eqnarray}
For the $I={2}$ $\pi \pi$ scattering length that we compute here, the
difference between the exact solution and the approximate solution in
eq.~(\ref{luscher_a}) is $\lsim 1\%$. For the volumes we
consider (with $L\simeq 2.5~{\rm fm}$), the center-of-mass momentum
is obviously non-zero and therefore one should keep in mind
the effective range expansion:
\begin{eqnarray}
p \cot\delta(p) & = & \frac{1}{a}\ + \ \frac{1}{2}\,r\,p^2 \ +\ {\cal O} (p^4)
\ \ \ ,
\label{eq:ERTa}
\end{eqnarray}
where $r$ is the effective range, which appears at ${\cal
  O}\left({1/L^6}\right)$ in eq.~(\ref{luscher_a}), and include the truncation
of eq.~(\ref{eq:ERTa}) as a source of systematic uncertainty.

%%%%%%%%%%%%%%%%%%%%%%%%%%%%%%%%%%%%
\section{Data Analysis and Chiral and Continuum Extrapolation}
\label{sec:Extrapolate}

\subsection{Results of the Lattice Calculation}
\label{sec:ResultsA}

\begin{figure}[!ht]
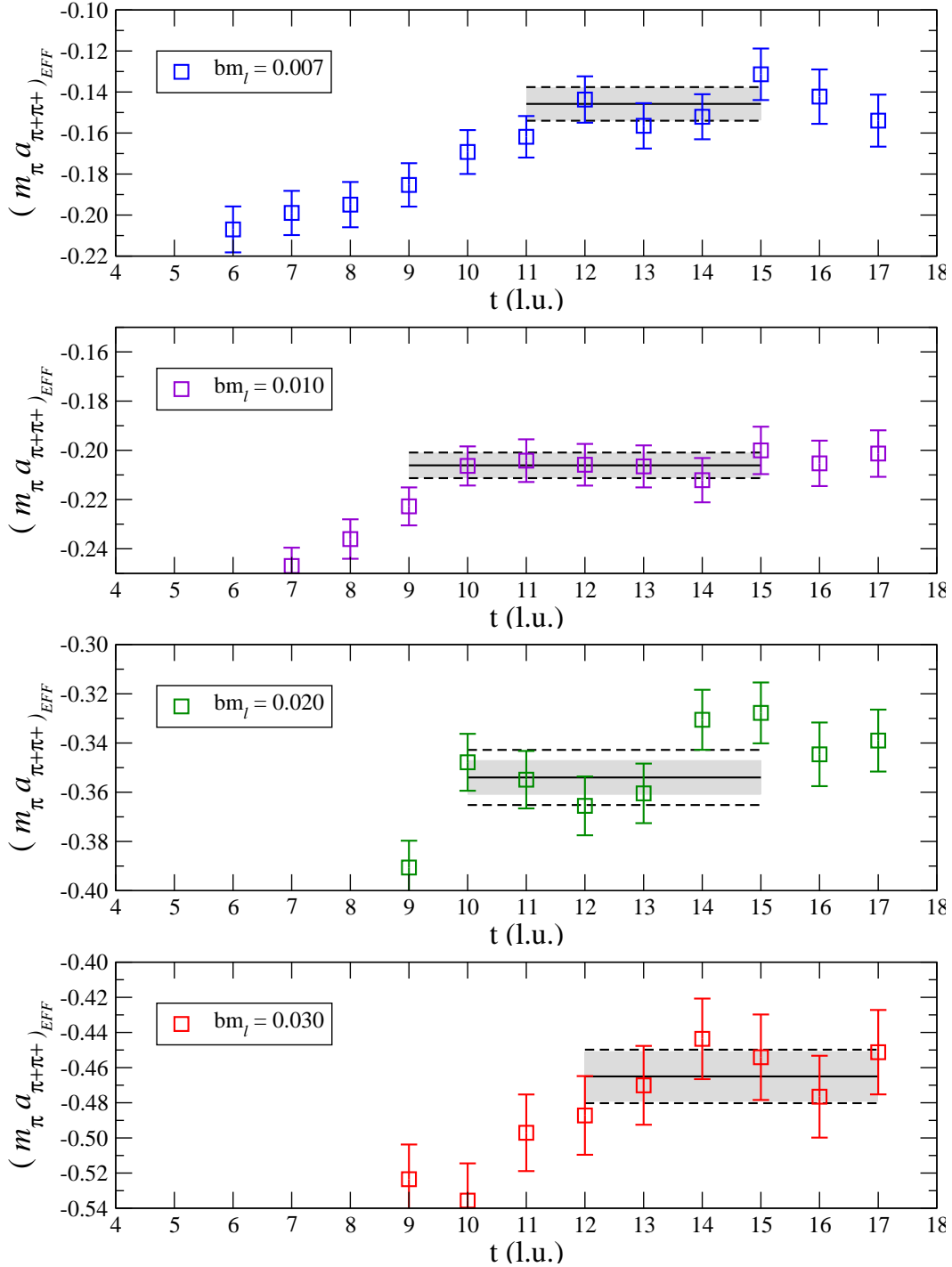

\centering                  
\includegraphics*[width=0.85\textwidth,viewport=2 5 700 240,clip]{EFFmpia007.eps}
\hfill
\includegraphics*[width=0.85\textwidth,viewport=2 5 700 240,clip]{EFFmpia010.eps}
\hfill
\includegraphics*[width=0.85\textwidth,viewport=2 5 700 240,clip]{EFFmpia020.eps}
\hfill
\includegraphics*[width=0.85\textwidth,viewport=2 5 700 240,clip]{EFFmpia030.eps}
\caption{\it 
The effective $\pi^+\pi^+$ scattering length times the effective $\pi$ mass
as a function of time slice arising from smeared sinks. The solid black lines and
shaded regions are fits with 1-$\sigma$ statistical 
uncertainties tabulated in Table~\ref{table:su2fits}.
The dashed lines are estimates of the systematic uncertainty due to fitting, also given
in Table~\ref{table:su2fits}.}
\label{fig:SSPPplots}
\end{figure}

\noindent It is convenient to present the results of our lattice
calculation in ``effective scattering length'' plots, simple variants
of effective-mass plots.  The effective energy splitting is formed
from the ratio of correlation functions
\begin{eqnarray}
\Delta E_{\pi^+ \pi^+}(t) & = & \log\left({ G_{\pi^+ \pi^+}(0,t)\over  G_{\pi^+ \pi^+}(0,t+1)}\right)
\  \ ,
\label{eq:effene} 
\end{eqnarray}
which in the limit of an infinite number of gauge configurations would
become a constant at large times that is equal to the lowest energy of
the interacting $\pi^+$'s in the volume.  At each time-slice, $\Delta
E_{\pi^+ \pi^+}(t)$ is inserted into eq.~(\ref{eq:energies}) (or
eq.~(\ref{luscher_a})), to give a scattering length at each time
slice, $a_{\pi^+\pi^+}(t)$.  It is customary to consider the
dimensionless quantity given by the pion mass times the scattering
length, $m_\pi\; a_{\pi^+\pi^+}$, where $m_\pi(t)$ is the pion
effective mass, in order to remove scale-setting uncertainties.  For
each of the MILC ensembles that we analyze, the effective scattering
lengths are shown in fig.~\ref{fig:SSPPplots}.  The statistical uncertainty
at each time slice has been generated with the Jackknife procedure.
The values of the pion masses, decay constants and $\pi\pi$ energy-shifts that we have
calculated are shown in Table~\ref{table:su2fits}.
\begin{table}[ht]
\caption{The summary table of raw fit quantities required for the two-flavor analysis.
The first uncertainties are statistical, the second uncertainties are systematic uncertainties due to fitting
and the third uncertainty, when present, is a comprehensive systematic uncertainty, as discussed in the text.}
\label{table:su2fits}
\resizebox{!}{2.58cm}{
\begin{tabular}{@{}|c | c | c | c | c |}
\hline
\  Quantity \ & 
\ \ \ $\qquad m_l=0.007\qquad$\ \ \ & 
\ \ \ $\qquad m_l=0.010\qquad$\ \ \ & 
\ \ \ $\qquad m_l=0.020\qquad$\ \ \ & 
\ \ \ $\qquad m_l=0.030\qquad$\ \  \  \\
\hline
\ Fit Range  \ & 
$\qquad 8-12\qquad$ & 
$\qquad 8-13 \qquad$ & 
$\qquad 7-13\qquad$  & 
$\qquad 9-12\qquad$ \\
\hline
$m_\pi$ (l.u.) & $0.18454(58)(51)$ & 
        $0.22294(31)(09)$ & 
        $0.31132(28)(21)$ & 
        $0.37407(49)(12)$\\
$f_\pi$ (l.u.) & $0.09273(29)(42)$ & 
        $0.09597(16)(10)$ & 
        $0.10179(12)(28)$ & 
        $0.10759(28)(17)$\\
$m_\pi/f_\pi$ & $1.990(11)(14)$ & 
        $2.3230(57)(30)$ & 
        $3.0585(49)(95)$ & 
        $3.4758(98)(60)$\\
\hline
\ Fit Range  \ & 
$\qquad 11-15\qquad$ & 
$\qquad 9-15 \qquad$& 
$\qquad 10-15\qquad$  & 
$\qquad 12-17\qquad$  \\
\hline
$\Delta E_{\pi\pi}$ (l.u.) & $0.00779(47)(14)$  & 
        $0.00745(20)(07)$ & 
        $0.00678(18)(20)$ & 
        $0.00627(23)(10)$\\
$m_\pi a_{\pi\pi}^{I=2}$ ($b\ne 0$) & 
        $-0.1458(78)(25)(14)$ & 
        $-0.2061(49)(17)(20)$ & 
        $-0.3540(68)(89)(35)$  & 
        $-0.465(14)(06)(05)$\\
$l^{I=2}_{\pi\pi}$\ ($b\ne 0$) & 
        $6.1(1.9)(0.7)(0.4)$ & 
        $5.23(68)(24)(28)$ & 
        $6.53(32)(42)(16)$ &
        $6.90(40)(18)(13)$\\
$\delta\ (b\ne 0) ({\rm degrees}) $ 
& $-1.71(14)(04)$
& $-2.181(81)(28)$
& $-3.01(09)(12)$
& $-3.46(17)(07)$\\
$|{\bf p}|/m_\pi$ 
& $0.2032(60)(18)$
& $0.1836(25)(09)$
& $0.1480(17)(23)$
& $0.1298(24)(10)$\\
\hline
\end{tabular}}
\end{table}
\begin{table}[ht]
\caption{Summary table for fit quantities extrapolated to the continuum with two-flavor MA$\chi$PT. 
The first row corresponds to the overall mixed action correction to the scattering
length. The uncertainties are discussed in detail in Section~\ref{sec:Systerrors}.
The second and third rows are the continuum limit scattering length and low-energy constant.
The first uncertainties are statistical and the second uncertainties are comprehensive systematic uncertainties.}
\label{table:su2fitspart2}
\resizebox{!}{1.02cm}{
\begin{tabular}{@{}|c | c | c | c | c |}
\hline
\  Quantity \ & 
\ \ \ $\qquad m_l=0.007\qquad$\ \ \ & 
\ \ \ $\qquad m_l=0.010\qquad$\ \ \ & 
\ \ \ $\qquad m_l=0.020\qquad$\ \ \ & 
\ \ \ $\qquad m_l=0.030\qquad$\ \  \  \\
\hline
$\Delta\left(m_\pi a_{\pi\pi}^{I=2}\right)$ 
& 0.0033(02)(02)(32)(55) 
& 0.0030(02)(04)(35)(22) 
& 0.0023(01)(10)(36)(03)
& 0.0018(01)(16)(32)(01) \\
\hline
$m_\pi a_{\pi\pi}^{I=2}$ ($b\rightarrow 0$) & 
        $-0.1491(78)(32)$ & 
        $-0.2091(49)(34)$ & 
        $-0.356(07)(11)$  & 
        $-0.467(14)(09)$\\
$l^{I=2}_{\pi\pi}$ ($b\rightarrow 0$) & 
        $5.3(1.9)(1.8)$ & 
        $4.83(68)(73)$ & 
        $6.42(32)(51)$ &
        $6.85(40)(27)$\\
\hline
\end{tabular}}
\end{table}
%

%%%%%%%%%%%%%%%%%%%%%%%%%%%%%%%%%5
\subsection{Two-Flavor Mixed-Action $\chi$-PT at One Loop}
\label{sec:ResultsB}

\noindent The mixed-action corrections for the $I=2\ \pi\pi$
scattering length have been determined in Ref.~\cite{Chen:2005ab}.  It
was demonstrated that when the extrapolation formulae for this system
are expressed in terms of the lattice-physical parameters~\footnote{
We denote quantities that are computed directly from the correlation functions, 
such as $m_\pi$, as lattice-physical quantities. These
are not extrapolated to the continuum, to infinite-volume or to the
physical point. }  as computed on the lattice, $m_\pi$, and
$f_\pi$, there are no lattice-spacing-dependent counterterms at
$\mathcal{O}(b^2)$, $\mathcal{O}(b^4)$ or $\mathcal{O}(m_\pi^2 b^2)
\sim \mathcal{O}(b^4)$. This was explained to be a general feature of
the two-meson systems at this order, including the non-zero momentum
states~\cite{Chen:2006wf}.  There are additional lattice-spacing
corrections due to the hairpin interactions present in mixed-action
theories, but for our scheme of domain-wall valence propagators
calculated in the background of the asqtad improved MILC gauge
configurations, these contributions are completely calculable without
additional counterterms at NLO, as they depend only upon valence meson
masses and the staggered taste-identity meson mass
splitting~\cite{Chen:2005ab,Chen:2006wf} which has been
computed~\cite{Aubin:2004fs}.  This allows us to precisely determine
the predicted mixed-action corrections for the scattering lengths at
the various pion masses used in this work.  In two-flavor MA$\chi$-PT
(i.e. including finite lattice-spacing corrections) the chiral
expansion of the scattering length at NLO takes the
form~\cite{Chen:2006wf}
\begin{eqnarray}
m_\pi\ a_{\pi\pi}^{I=2}(b\ne 0)  =  
-{m_\pi^2\over 8\pi f_\pi^2}
 \Biggl\{
1 + {m_\pi^2\over 16\pi^2 f_\pi^2}\ \Biggl[
3 \log\left({m_\pi^2\over\mu^2}\right)\ -\ 1\ -\ l_{\pi\pi}^{I=2}(\mu)\ -\
{\tilde\Delta_{ju}^4\over 6 m_\pi^4}\ \Biggr]
\ \Biggr\} \ ,
\label{eq:su2chiPT}
\end{eqnarray}
where it is understood that $m_\pi$ and $f_\pi$ are the lattice-physical parameters~\cite{Chen:2006wf}
and
\begin{eqnarray}
\tilde{\Delta}_{ju}^2 &\equiv \tilde{m}_{jj}^2 - m_{uu}^2
                = 2 B_0 (m_j- m_u) + b^2 \Delta_I +\dots\, ,
\label{eq:su2chiPTB}
\end{eqnarray}
where $u$ denotes a valence quark and $j$ denotes a sea-quark, and we
are using isospin-symmetric sea and valence quarks. $\tilde{m}_{jj}$
($m_{uu}$) is the mass of a meson composed of two sea (valence) quarks
of mass $m_j$ ($m_u$) and the dots denote higher-order corrections to
the meson masses.  Clearly eq.~(\ref{eq:su2chiPT}), which contains all
$\mathcal{O}(m_\pi^2 b^2)$ and $\mathcal{O}(b^4)$ lattice artifacts,
reduces to the continuum expression for the scattering
length~\cite{Gasser:1983yg} in the QCD limit where
$\tilde{\Delta}_{ju}^2\rightarrow 0$~\footnote{The counterterm
$l_{\pi\pi}^{I=2}(\mu)$ is, of course, the same counterterm that
appears in continuum $\chi$PT.}.  It is worth noting that
eq.~(\ref{eq:su2chiPT}), and the subsequent expression for the
three-flavor theory, become the partially-quenched formulae in the
continuum limit.  
Therefore, they are the correct extrapolation
formulae to use in the case of non-degenerate valence- and
sea-quark masses, as is implied by
eq.~(\ref{eq:su2chiPT}) and eq.~(\ref{eq:su2chiPTB}). This modification of the partially-quenched
formulae can be understood on more general grounds, as mixed-action
theories with chirally-symmetric valence fermions exhibit many
universal features~\cite{Chen:2007ug}.

With domain-wall fermion masses tuned to match the
staggered Goldstone pion~\cite{Renner:2004ck,Edwards:2005kw}, one
finds $\tilde{\Delta}_{ju}^2\ =\ b^2 \Delta_I$.  The various fit
parameters relevant to the two-flavor extrapolation are presented in
Table~\ref{table:su2fits}.  For each ensemble we determine $m_\pi\
a_{\pi\pi}^{I=2}$, and then use the chiral extrapolation formula to
extract a value of the counterterm $l_{\pi\pi}^{I=2}(\mu=f_\pi)$, with
a statistical uncertainty determined with the Jackknife procedure.
The systematic uncertainties are propagated through in quadrature. 
The results of the two-flavor extrapolation to the continuum are shown in
Table~\ref{table:su2fitspart2}.

Fitting to lattice data at the lightest accessible values of the quark
masses will optimize the convergence of the chiral expansion.  While
we only have four different quark masses in our data set, with pion
masses, $m_\pi\sim 290~{\rm MeV}$, $350~{\rm MeV}$, $490~{\rm
  MeV}$ and $590~{\rm MeV}$,
fitting all four data sets and then ``pruning'' the heaviest data set
and refitting provides a useful measure of the convergence of the
chiral expansion. Hence, in ``fit A'', we fit the
$l_{\pi\pi}^{I=2}(\mu=f_\pi)$'s extracted 
from all four lattice ensembles (m007, m010, m020 and m030) to a constant, 
while in ``fit B'', we
fit the $l_{\pi\pi}^{I=2}(\mu=f_\pi)$'s from the lightest three lattice ensembles (m007, m010 and
m020). In ``fit C'', we fit the $l_{\pi\pi}^{I=2}(\mu=f_\pi)$'s from the lightest two lattice
ensembles (m007 and m010). Results are given in Table~\ref{tab:FitResultsNLOsu2}.
\begin{table}[!ht]
 \caption{Results of the fits in two-flavor Mixed-Action $\chi$-PT.
The values of $m_\pi\ a_{\pi\pi}^{I=2}$ correspond to the extrapolated values at the physical point.
The first uncertainty is statistical and the second is a comprehensive systematic uncertainty.}
\label{tab:FitResultsNLOsu2}
\begin{ruledtabular}
\begin{tabular}{cccc}
FIT &  $l_{\pi\pi}^{I=2}(\mu=f_\pi)$ &  $m_\pi\ a_{\pi\pi}^{I=2}$
(extrapolated) & $\chi^2$/dof \\
\hline
A & $6.43\pm 0.23\pm 0.26$  &  $-0.043068\pm 0.000076\pm 0.000085$  & $1.17$  \\
B & $5.97\pm 0.29\pm 0.42$  &  $-0.043218\pm 0.00009\pm 0.00014$  & $0.965$  \\
C & $4.89\pm 0.64\pm 0.68$  &  $-0.04357\pm 0.00021\pm 0.00022$  & $0.054$  \\
\end{tabular}
\end{ruledtabular}
\end{table}

Taking the range of parameters spanned by fits A-C
one finds:
\begin{eqnarray}
l_{\pi\pi}^{I=2}(\mu=f_\pi) & = & 5.4\pm 1.4
\nonumber\\
m_\pi\ a_{\pi\pi}^{I=2} & = & -0.04341\pm 0.00046 \ .
\end{eqnarray}

\begin{figure}[!ht]
\vskip 0.95cm
\centering                  
\centerline{{\epsfxsize=6.in \epsfbox{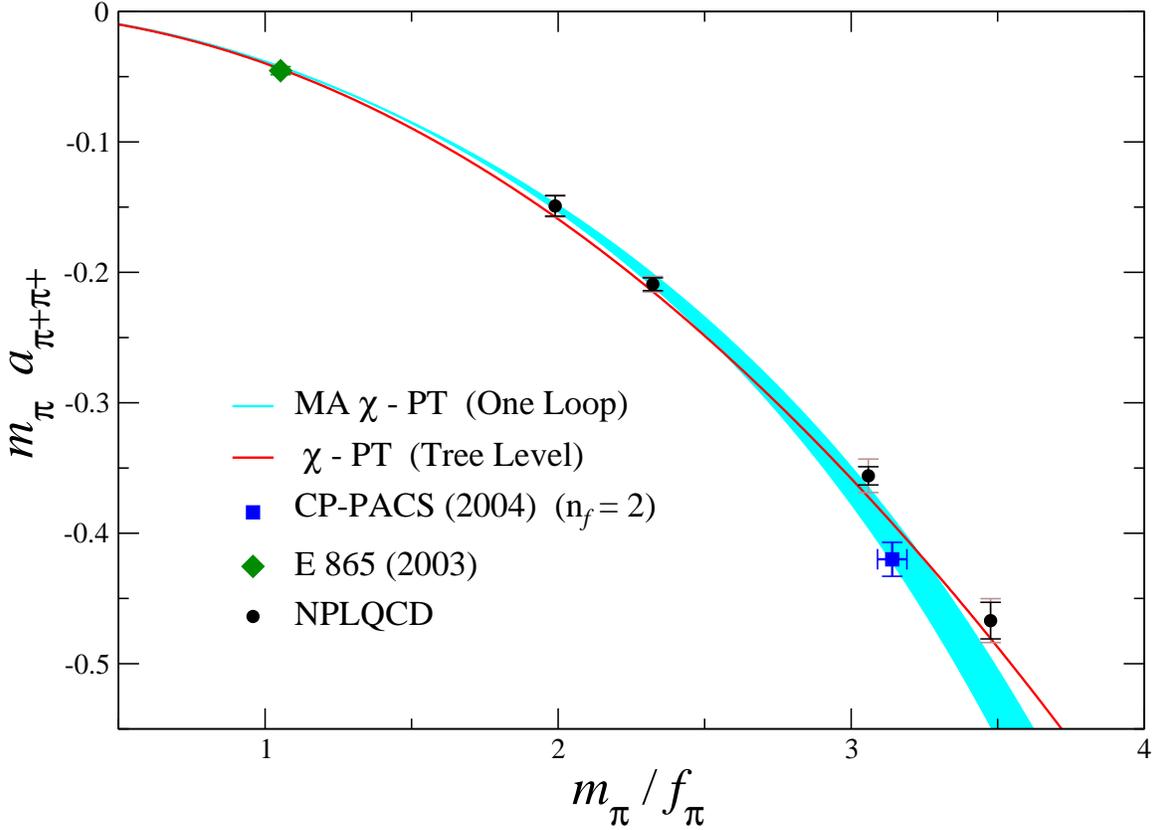}}}
\caption{
$m_\pi \ a_{\pi\pi}^{I=2}$ vs. $m_\pi/f_\pi$ (ovals) with statistical (dark bars) and
systematic (light bars) uncertainties.  Also shown are the
experimental value from Ref.~\cite{Pislak:2003sv} (diamond) and the
lowest quark mass result of the $n_f=2$ dynamical calculation of
CP-PACS~\cite{Yamazaki:2004qb} (square).  The blue band corresponds to
a weighted fit to the lightest three data points (fit B) using the one-loop MA$\chi$-PT
formula in eq.~(\ref{eq:su2chiPT}) (the shaded region corresponds only
to the statistical uncertainty). The red line is the tree-level $\chi$-PT
result.
The experimental data is not used in the chiral extrapolation fits.
}
\label{fig:CAplot}
\end{figure}

In Fig.~\ref{fig:CAplot} we show the results of our calculation, along
with the lowest mass $n_f=2$ point from CP-PACS (not included in our
fit).  We also show the tree-level prediction and the results of our
two-flavor fit described in this section.  The experimental point
shown in Fig.~\ref{fig:CAplot} is not included in the fit and
extrapolation.  It is interesting that the lattice data indicates
little deviation from the tree level $\chi$PT curve. The significant
deviation of the extrapolated scattering length from the tree-level
result is entirely a consequence of fitting to MA$\chi$PT at one-loop
level.

%

%%%%%%%%%%%%%%%%%%%%%%%%%%%%%%%%%%%%%%
\subsection{Three-Flavor Mixed-Action $\chi$-PT at One Loop}
\label{sec:ResultsC}

\noindent An important check of the systematic uncertainties involved in the
chiral extrapolation is to perform the same analysis using
three-flavor MA$\chi$-PT~\cite{Chen:2005ab,Chen:2006wf} as both the
real world and our lattice calculation have three active light
flavors.  In addition to the computations presented in
Table~\ref{table:su2fits}, it is necessary to determine masses and
decay constants for the kaon and the $\eta$.  We use the
Gell-Mann--Okubo mass-relation among the mesons to determine the
$\eta$ mass, which we do not compute in this lattice calculation due
the enormous computer resources (beyond what is available to us)
required to compute the disconnected contributions. This procedure is
consistent to the order in the chiral expansion to which we are working.
\begin{table}[ht]
\caption{The summary table of quantities required for the three-flavor analysis.
A `` $^*$ '' denotes that the Gell-Mann-Okubo mass relation among the mesons has
been used to determine this quantity. The first uncertainties are statistical and the
second are systematic (that are discussed in the text).}
\label{table:su3fits}
\resizebox{!}{2.3cm}{
\begin{tabular}{@{}|c | c | c | c | c |}
\hline
\  Quantity \ & 
\ \ \ $\qquad m_l=0.007\qquad$ \ \ \ & 
\ \ \ $\qquad m_l=0.010\qquad$ \ \ \ & 
\ \ \ $\qquad m_l=0.020\qquad$ \ \ \ & 
\ \ \ $\qquad m_l=0.030\qquad$  \ \ \ \\
\hline
\ Fit Range  \ & 
$\qquad 8-14\qquad$ & 
$\qquad 9-14 \qquad$ & 
$\qquad 9-13\qquad$  & 
$\qquad 9-13\qquad$ \\
\hline
$m_K$ (l.u.) & $0.36839(40)(29)$ & 
        $0.37797(30)(03)$ & 
        $0.40540(31)(32)$ & 
        $0.42976(41)(20)$\\
$m_\eta$ (l.u.) $^*$ & $0.41182(43)(36)$ & 
        $0.41703(32)(04)$ & 
        $0.43224(33)(46)$ & 
        $0.44688(38)(26)$\\
$m_\eta/f_\pi$ $^*$ & $4.447(19)(20)$ & 
        $4.3517(96)(43)$ & 
        $4.246(06)(12)$ & 
        $4.154(11)(05)$\\
$\tilde m_X/f_\pi$ $^*$ & $5.408(23)(24)$ & 
        $5.271(11)(05)$ & 
        $5.087(07)(14)$ & 
        $4.927(13)(06)$\\
\hline
$\Sigma$ $^*$ & 
        $-0.0015(01)$ & 
        $-0.0027(00)$ & 
        $-0.0079(01)$ & 
        $-0.0130(03)$\\
$\Gamma$ $^*$ & 
        $0.0011(01)$ & 
        $0.0003(01)$ & 
        $-0.0012(01)$ & 
        $-0.0018(01)$\\
$m_\pi a_{\pi\pi}^{I=2}$ ($b\rightarrow 0$) $^*$ & 
$-0.1470(78)(70)$ &
$-0.2065(49)(50)$ &
$-0.353(07)(10)$ & 
$-0.462(14)(08) $ \\
$ 32 (4\pi)^2 L_{\pi\pi}^{I=2} $ $^*$ &
$6.4(1.9)(1.7)$ &
$5.66(67)(68)$ &
$7.07(32)(48)$ & 
$7.44(40)(21)$  \\
\hline
\end{tabular}}
\end{table}

The chiral expansion of the $\pi^+\pi^+$ scattering length in three-flavor mixed-action 
$\chi$PT takes the form~\cite{Chen:2006wf}:
\begin{eqnarray}
m_\pi\ a_{\pi\pi}^{I=2}(b\ne 0) & = & 
-{m_\pi^2\over 8\pi f_\pi^2}
 \left\{ 
1 + {m_\pi^2\over 16\pi^2 f_\pi^2} \left[ 
3 \log\left({m_\pi^2\over\mu^2}\right) - 
32 (4\pi)^2\ L_{\pi\pi}^{I=2}(\mu) 
 + {1\over 9}\log\left({\tilde m_X^2\over\mu^2}\right)
 - {8\over 9}
\right.\right.\nonumber\\
&& \left.\left. \qquad \qquad \qquad \qquad \qquad 
\ -\
{\tilde\Delta_{ju}^4\over 6 m_\pi^4}
\ +\ \sum_{n=1}^4\ \left({\tilde\Delta_{ju}^2\over m_\pi^2}\right)^n\ {\cal
F}_n\left({m_\pi^2\over\tilde m_X^2}\right)
\right]
\ \right\}
\ \ \ ,
\label{eq:su3chiPT}
\end{eqnarray}
where $\tilde m_X^2  =  m_\eta^2 + b^2 \Delta_I$, and
\begin{eqnarray}
        {\cal F}_1(y) &=& -\frac{2y}{9(1-y)^2} \Big[ 5(1-y) +(3 +2y)\ln (y) \Big],  \nonumber\\
        {\cal F}_2(y) &=& \frac{2y}{3(1-y)^3} \Big[ (1-y)(1+3y) +y(3 +y)\ln(y) \Big], \nonumber\\
        {\cal F}_3(y) &=& \frac{y}{9(1-y)^4} \Big[ (1-y) (1 -7y -12y^2) -2y^2(7+2y) \ln(y) \Big], \nonumber\\
        {\cal F}_4(y) &=& -\frac{y^2}{54 (1-y)^5} \Big[ (1-y) (1 -8y -17y^2) -6y^2(3+y)\ln(y) \Big]\, .
\label{eq:coolFs}
\end{eqnarray}
In addition, it is useful to define the quantities:
\begin{eqnarray}
\Gamma \equiv -{2\pi m_\pi^4\over (4\pi f_\pi)^4} \left[
-{\tilde\Delta_{ju}^4\over 6 m_\pi^4}+
\sum_{n=1}^4  \left({\tilde\Delta_{ju}^2\over m_\pi^2}\right)^n\ 
{\cal F}_n\left({m_\pi^2\over\tilde m_X^2}\right)
\right]
\label{eq:gam}
\end{eqnarray}
and
\begin{eqnarray}
\Sigma \equiv -{m_\pi^2\over 8\pi f_\pi^2}{m_\pi^2\over 16\pi^2 f_\pi^2} {1\over
  9}\log\left({\tilde m_X^2\over f_\pi^2}\right) \ ,
\label{eq:sig}
\end{eqnarray}
whose numerical values for the various ensembles are given in Table~\ref{table:su3fits}.

For the three-flavor analysis,
we follow the same procedure of ``pruning'' the data as in the
two-flavor analysis, giving the  results shown in Table~\ref{tab:FitResultsNLOsu3}.
\begin{table}[!ht]
 \caption{Results of the NLO fits in three-flavor Mixed-Action $\chi$-PT.
The values of $m_\pi\ a_{\pi\pi}^{I=2}$ correspond to the extrapolated values at the physical point.
The first uncertainty is statistical and the second is a comprehensive systematic uncertainty.}
\label{tab:FitResultsNLOsu3}
\begin{ruledtabular}
\begin{tabular}{cccc}
FIT &  $32(4\pi)L_{\pi\pi}^{I=2}(\mu=f_\pi)$ &  $m_\pi\ a_{\pi\pi}^{I=2}$
(extrapolated) & $\chi^2$/dof \\
\hline
D & $7.09\pm 0.23\pm 0.23$  &  $-0.042992\pm 0.000076\pm 0.000077$  & $0.969$  \\
E & $6.69\pm 0.29\pm 0.39$  &  $-0.04312\pm 0.00009\pm 0.00013$  & $0.803$  \\
F & $5.75\pm 0.63\pm 0.64$  &  $-0.04343\pm 0.00021\pm 0.00021$  & $0.073$  \\
\end{tabular}
\end{ruledtabular}
\end{table}
Taking the range of parameters spanned by fits D-F
one finds:
\begin{eqnarray}
32(4\pi)L_{\pi\pi}^{I=2}(\mu=f_\pi) & = & 6.2\pm 1.2
\nonumber\\
m_\pi\ a_{\pi\pi}^{I=2} & = & -0.04330\pm 0.00042
\ \ \ .
\end{eqnarray}

%%%%%%%%%%%%%%%%%%%%%%%%%%%%%%%%%%%%%%%%%%%%%%%%%%%%%%%%%%%%%
\section{Systematic Uncertainties}
\label{sec:Systerrors}

\noindent This section describes the sources of systematic uncertainty that need
to be quantified.

%%%%%%%%%%%%%%%%%%%%%%%%%%%%%%%%
\subsection{Higher-Order Effects in Mixed-Action $\chi$-PT}

\begin{table}[t]
\caption{\label{tab:pipi_2_errors} Corrections and uncertainties in  $m_\pi
  a_{\pi\pi}^{I=2}$ for $n_f=2$.}
\begin{tabular}{|c|c|c|c|c|}
\hline
Quantity &  \ \ \ \  $m_l = 0.007$ \ \ \ \   & \ \ \ \   $m_l = 0.010$ \ \ \ \   & \ \ \ \   $m_l = 0.020$ \ \ \ \   & \ \ \ \   $m_l = 0.030$ \ \ \ \   \\ \hline
$\Delta_{MA} \left(m_\pi a_{\pi\pi}^{I=2} \right)$ 
& 0.0033(02)(02) & 0.0030(02)(04) & 0.0023(01)(10) & 0.0018(01)(16) \\
$\Delta_{FV} \left(m_\pi a_{\pi\pi}^{I=2} \right)$ 
& $\pm 0.0055$ & $\pm 0.0022$ & $\pm 0.0003$ & $\pm 0.0001$ \\
$\Delta_{m_{res}} \left(m_\pi a_{\pi\pi}^{I=2} \right)$ 
& $\pm 0.0032$ & $\pm 0.0035$ & $\pm 0.0036$ & $\pm 0.0032$ \\
\hline
\end{tabular}
\end{table}
\begin{table}[t]
\caption{\label{tab:pipi_2p1_errors} Corrections and uncertainties in  $m_\pi
  a_{\pi\pi}^{I=2}$ for $n_f=2+1$.}
\begin{tabular}{|c|c|c|c|c|}
\hline 
Quantity & \ \ \ \ $m_l = 0.007$ \ \ \ \ & \ \ \ \ $m_l = 0.010$\ \ \ \  & \ \ \ \ $m_l = 0.020$\ \ \ \  & \ \ \ \ $m_l = 0.030$ \ \ \ \  \\ \hline
$\Delta_{MA} \left(m_\pi a_{\pi\pi}^{I=2} \right)$ 
& 0.0012(01)(02) & 0.0004(01)(04) & -0.0015(03)(10) & -0.0027(05)(16) \\
$\Delta_{FV} \left(m_\pi a_{\pi\pi}^{I=2} \right)$ 
& $\pm0.0024$ & $\pm0.0005$ & $\pm0.0001$ & $\pm0.00006$ \\
$\Delta_{m_{res}} \left(m_\pi a_{\pi\pi}^{I=2} \right)$ 
& $\pm 0.0032$ & $\pm 0.0035$ & $\pm 0.0036$ & $\pm 0.0032$ \\
\hline
\end{tabular}
\end{table}

\noindent 
We rely on the power counting associated with the chiral expansion of
the Mixed-Action $\chi$PT to estimate the size of the
lattice-spacing artifacts arising at $\mathcal{O}(m_\pi^4 b^2)$.  To
be conservative, we have estimated these corrections to be of the
general size
\begin{equation}
\mathcal{O}(m_\pi^4 b^2) \        
\sim\  \frac{2\pi m_\pi^4}{(4\pi f_\pi)^4} \frac{b^2 \Delta_\mathrm{I}}{(4\pi
          f_\pi)^2}
\ \ \  .
\label{eq:maerr}
\end{equation}
We treat these estimates as uncertainties in the predicted NLO MA$\chi$PT
corrections which can be determined from eq.~(\ref{eq:su2chiPT})
and eq.~(\ref{eq:su3chiPT}). We provide these predicted corrections and their
uncertainties in the form
\begin{equation}
        \Delta_{MA} \left( m_\pi a_{\pi\pi}^{I=2} \right) 
\ = \  m_\pi a_{\pi\pi}^{I=2}  \Big|_{MA} \ -\   m_\pi a_{\pi\pi}^{I=2}  \Big|_{\chi
  PT}
\ \ \ \ .
\end{equation}
The values of these corrections are shown in
Tables~\ref{tab:pipi_2_errors} and \ref{tab:pipi_2p1_errors}.  The
first uncertainty in these corrections is statistical and is 
associated with the meson masses, decay constants and the
taste-identity mass splitting, $b^2 \Delta_\mathrm{I}$.  
The second uncertainty is the power counting estimate of the higher-order corrections
of $\mathcal{O}(m_\pi^4 b^2)$ as estimated in eq.~(\ref{eq:maerr}).  
The calculable
corrections to $m_\pi a_{\pi\pi}^{I=2}$ at
$\mathcal{O}(m_\pi^2 b^2, b^4)$ 
are $2.3\%$, $1.5\%$, $0.65\%$ and $0.39\%$ effects for the 007, 010, 020 and 030 ensembles,
respectively, from which we conclude that the $\mathcal{O}(m_\pi^4 b^2)$
contributions are significantly less than $\sim 1\%$.

%%%%%%%%%%%%%%%%%%%%%%%%%%%%%%%%
\subsection{Finite-Volume Effects in Mixed-Action $\chi$-PT}

\noindent 
The universal relation between the two-particle energy levels in a
finite volume and their infinite-volume scattering parameters receives
non-universal corrections which are exponentially suppressed by the
lattice size and dominated by the lightest particle in the
spectrum. These scale generically as $e^{-m_\pi
L}$~\cite{Luscher:1985dn,Gasser:1987ah}.  In
Ref.~\cite{Bedaque:2006yi}, the leading exponential volume corrections
to $p\cot \delta(p)$ were determined in the $I=2\ \pi\pi$ scattering
channel in $\chi$PT.  However, in order to determine the leading
finite-volume corrections to this mixed-action calculation, hairpin
diagrams present in the mixed-action theory must also be included.
For the $I=2\ \pi\pi$ system, there are additional hairpin diagrams
present in the $t$ and $u$ channel scattering
diagrams~\cite{Chen:2005ab}.  The finite-volume corrections from these
diagrams are larger than those in continuum $\chi$PT, but are opposite
in sign and therefore the overall magnitude of the correction is
similar to that given in Ref.~\cite{Bedaque:2006yi}.  We note that as
these contributions vanish in the continuum limit, they are actually
finite-volume finite-lattice-spacing corrections, and not just
finite-volume corrections, and hence scale as $b^2 \exp(-m_\pi L)$ at small
lattice spacing.

As with the mixed-action lattice-spacing corrections, we denote these finite-volume modifications as
\begin{equation}
        \Delta_{FV} \left( m_\pi a_{\pi\pi}^{I=2}  \right) =
 m_\pi a_{\pi\pi}^{I=2}  \Big|_{FV} - m_\pi a_{\pi\pi}^{I=2}  \Big|_{\infty V}\, ,
\end{equation}
and they are shown in Tables~\ref{tab:pipi_2_errors} and
\ref{tab:pipi_2p1_errors}.  However, one should take note that the
effective-range contribution to $p \cot \delta(p)$, which behaves as a
power-law in the lattice size (and therefore is parametrically
enhanced over the exponential corrections) is not included in the
extraction of the scattering lengths.  While the exponential
modifications are numerically larger than our estimate of the
effective-range contributions at the light pion masses (see below),
the values of $\Delta_{FV}(m_\pi a_{\pi\pi}^{I=2})$ shown in
Tables~\ref{tab:pipi_2_errors} and \ref{tab:pipi_2p1_errors} are used
as estimates of the uncertainties due to higher-order finite-volume
effects.

%%%%%%%%%%%%%%%%%%%%%%%%%%%%%%%%
\subsection{Residual Chiral Symmetry Breaking}

\noindent 
The mixed-action formulae describing $\pi\pi$ scattering determined in
Refs.~\cite{Chen:2005ab,Chen:2006wf} have assumed that the valence
fermions have exact chiral symmetry, up to the quark-mass corrections.
The domain-wall propagators used in this work have a finite
fifth-dimensional extent and therefore residual chiral symmetry
breaking arising from the overlap of the left- and right-handed quark
fields bound to the opposite domain walls.  Due to the nature of this
residual chiral symmetry breaking in the domain-wall action, the
leading contributions can be parameterized as an additive shift to the
valence-quark masses~\cite{Shamir:1993zy,Furman:1994ky},
\begin{equation}
        m_l^{dwf} \rightarrow m_l^{dwf} + m_{res}\, .
\end{equation}
A full treatment of these effects involves three new spurion fields in
the effective field theory~\cite{Golterman:2004mf} but this is not
necessary for estimating the size of these contributions to the
$\pi\pi$ scattering lengths.  By expressing the calculated scattering
lengths and extrapolation formulae in terms of the lattice-physical
meson masses and decay constants, the dominant contributions from
residual chiral symmetry breaking are included, leaving
corrections at higher orders in the chiral expansion.  There will be
new operators similar to the Gasser-Leutwyler
operators~\cite{Gasser:1984gg} in the chiral Lagrangian, for example
\begin{eqnarray}\label{eq:m_res_ops}
        \bar{\mathcal{L}}& =&  
        2 B_0\, \bar{L}_4\; \textrm{str} \left( \partial_\mu \Sigma \partial^\mu \Sigma^\dagger \right) 
                \textrm{str} \left ( m_{res} \Sigma^\dagger + \Sigma m_{res}^\dagger \right) 
        \nonumber\\ 
        &&+ 8 B_0^2\,\bar{L}_6\; \textrm{str} \left( m_{q} \Sigma^\dagger + \Sigma m_{q}^\dagger \right)
                 \textrm{str} \left( m_{res} \Sigma^\dagger + \Sigma m_{res}^\dagger \right)
        +\dots
\end{eqnarray}
Naive dimensional analysis~\cite{Manohar:1983md} can be used to
estimate the size of the corrections due to these new operators, which
in the case of the $I=2\ \pi\pi$ system are given by
\begin{equation}\label{eq:m_res_error}
        \Delta_{m_{res}}(m_\pi a_{\pi\pi}^{I=2}) = \frac{8\pi m_\pi^4}{(4\pi f_\pi)^4}
                \frac{m_{res}}{m_l}\,  \ .
\end{equation}
There will be additional operators with two insertions of $m_{res}$ in
the place of $m_q$, but these are $\lsim 20\%$ of the uncertainty
already estimated for the residual chiral symmetry breaking. 
These uncertainties are denoted by
\begin{equation}
        \Delta_{m_{res}}(m_\pi a_{\pi\pi}^{I=2}) = m_\pi a_{\pi\pi}^{I=2} \Big|_{m_{res}} - m_\pi a_{\pi\pi}^{I=2}
        \Big|_{m_{res}=0}\, ,
\end{equation}
and are shown  in Tables~\ref{tab:pipi_2_errors} and \ref{tab:pipi_2p1_errors}.

%%%%%%%%%%%%%%%%%%%%%%%%%%%%%%%%
\subsection{Two Loops Effects}
\label{sec:ResultsD}

\noindent 
The two-loop expression for the scattering
length~\cite{Bijnens:1997vq,Colangelo:2001df} is given, in the continuum limit
of QCD, by
\begin{eqnarray}
m_\pi\ a_{\pi\pi}^{I=2}\ &=&\ -{m^2_\pi \over 8\pi f_\pi^2}\; \left\{ \ 
1\ +\ 
{m_\pi^2\over 16\pi^2 f_\pi^2}\; \left[\ 3 \log{\frac{m_\pi^2}{\mu^2}}
\ -1\ - \ l_{\pi\pi}^{I=2}(\mu )\ \right] 
\nonumber \right. \\
& & \left. \quad\qquad + \ {m_\pi^4\over 64\pi^4 f_\pi^4}\; \left[\ 
\frac{31}{6}\,\left(\log{\frac{m_\pi^2}{\mu^2}}\right)^2 \ +\ l^{(2)}_{\pi\pi}(\mu )\; 
\log{\frac{m_\pi^2}{\mu^2}}\ + \ l^{(3)}_{\pi\pi}(\mu )\ \right]
\right\},
\label{eq:ascattGLwithchexp2twoloop}
\end{eqnarray}
where $l^{(2)}_{\pi\pi}$ and $l^{(3)}_{\pi\pi}$ are linear combinations of undetermined
constants that appear in the $\mathcal{O}(p^4)$ and $\mathcal{O}(p^6)$  
chiral Lagrangians~\cite{Gasser:1983yg,Bijnens:1997vq}. Fitting all four data points allows 
for an extraction of the three counterterms with $\chi^2$/dof = 0.26. From the 
$68\%$ confidence-interval error ellipsoid we find an extrapolated value of:
\begin{eqnarray}
m_\pi\ a_{\pi\pi}^{I=2}& = & -0.0442\pm 0.0030 \ .
\end{eqnarray}
While it is gratifying to have a determination of the scattering length at two-loop level
that is consistent with the one-loop result, there are several caveats: i) the
two-loop expression in MA$\chi$-PT does not yet exist
and therefore the determination in eq.~(\ref{eq:ascattGLwithchexp2twoloop})
contains lattice-spacing artifacts at lower orders in
the chiral expansion than in the one-loop result; ii) This value is clearly
strongly dependent on the heaviest quark mass, which is, at best, at the boundary
of the range of validity of the chiral expansion. A reliable two-loop determination
will have to await further lattice data at quark masses closer to the chiral limit
than we currently possess.

%%%%%%%%%%%%%%%%%%%%%%%%%%%%%%%%
\subsection{Range Corrections}

\noindent 
It is straightforward to show that the range corrections enter at ${\cal
  O}\left(L^{-6}\right)$
in eq.~(\ref{luscher_a}).
Assuming that the effective range is of order the scattering length (the
scattering length is of natural size),
we expect a fractional uncertainty of $(m_\pi a)^2 p^2/2m_\pi^2$ due to
the omission of range corrections. For the ensembles that we
consider, this translates into an $0.5\%$ uncertainty in $m_\pi a_{\pi\pi}^{I=2}$. 
Allowing for the effective range to  exceed its natural value by a factor of
two, we assign a $1\%$ systematic uncertainty to $m_\pi a_{\pi\pi}^{I=2}$
determined on each ensemble.

%%%%%%%%%%%%%%%%%%%%%%%%%%%%%%%%
\subsection{Isospin Violation}

\noindent 
The calculation we have performed is in the limit of exact isospin
symmetry, as are the extrapolation formula we have used to analyze
the results.  The conventional discussion of the scattering length is in the
unphysical theory with $e=0$ and $m_u=m_d=m$, with
$m_\pi=m_{\pi^+}=139.57018\pm 0.00035~{\rm MeV}$ and $f_\pi=f_{\pi^+}
= 130.7\pm 0.14\pm 0.37~{\rm MeV}$.  Hence
$m_{\pi^+}/f_{\pi^+}=1.0679\pm 0.0032$, where the statistical and
systematic uncertainties have been combined in quadrature.  
We extrapolate the results of our lattice calculations to this value.

Unfortunately, we are presently unable to make precise predictions for
the real world in which isospin breaking occurs at the few-percent
level.  Extrapolation to the isospin-averaged pion mass (as opposed to
the charged pion mass), would introduce a shift of $\sim 2\%$ in
$m_\pi a_{\pi\pi}^{I=2}$.  This is larger than the uncertainty we have
determined at the charged pion mass.  It is clear that in order to
make predictions for real-world quantities at the $\sim 1\%$ level
from lattice QCD calculations, isospin-breaking and electromagnetism
will need to be incorporated into the lattice calculation.

%%%%%%%%%%%%%%%%%%%%%%%%%%%%%%%%%%%%%%%%%%%%%%%%%
\section{Discussion}
\label{sec:Conclude}

\begin{figure}[!ht]
\vskip 0.5in
\centerline{{\epsfxsize=4in \epsfbox{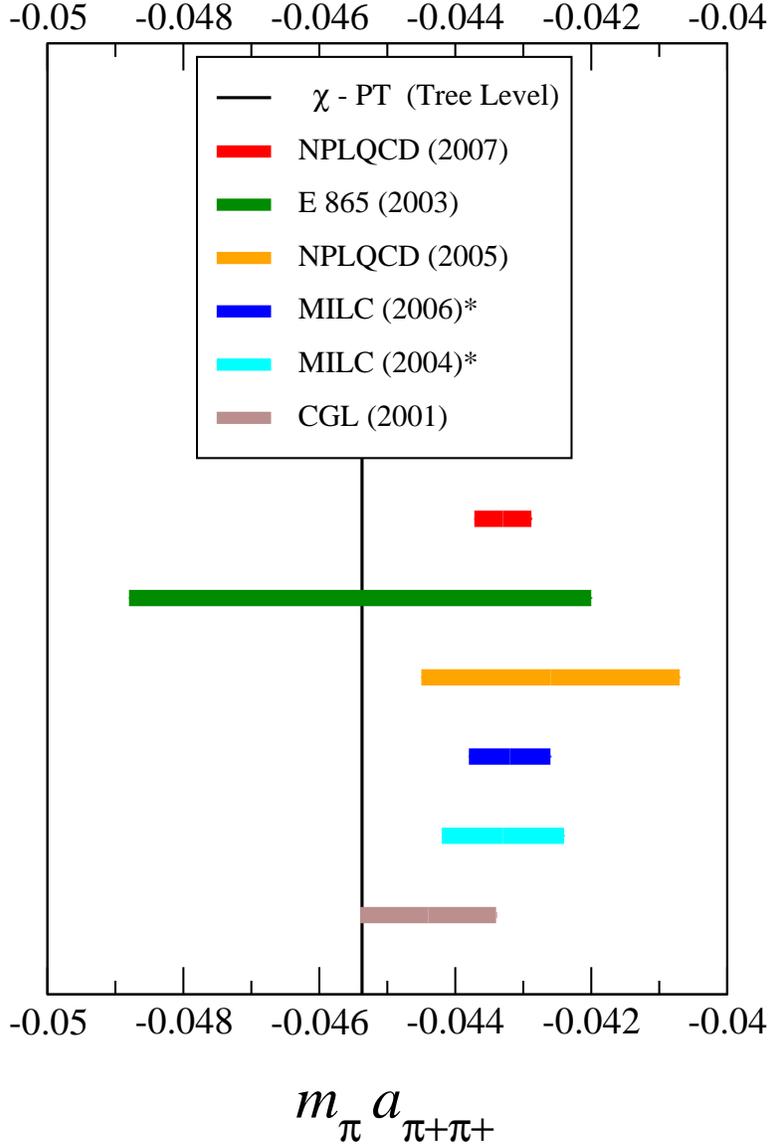}}}
\caption{ Bar chart of the various determinations of 
the $I=2$ $\pi\pi$ scattering length tabulated in
Table~\ref{tab:vardet}. We reiterate that 
the stars on the MILC results indicate that these are not lattice calculations of the
$I=2$ $\pi\pi$ scattering length but rather a hybrid prediction which
uses MILC's determination of various low-energy constants together
with the Roy equations.}
\label{fig:barchart}
\end{figure}

\noindent We have presented results of a lattice QCD calculation of
the $I=2$ $\pi\pi$ scattering length performed with domain-wall
valence quarks on asqtad-improved MILC configurations with 2+1
dynamical staggered quarks.  The calculations were performed at a
single lattice spacing of $b\sim 0.125~{\rm fm}$ and at a single
lattice spatial size of $L\sim 2.5~{\rm fm}$ with four values of the
light quark masses, corresponding to pion masses of $m_\pi\sim 290,
350$, $490~{\rm MeV}$ and $590~{\rm MeV}$. High statistics were
generated by computing up to twenty-four propagators per MILC configuration
at spatially- and temporally-displaced sources. We used one-loop
MA$\chi$-PT with two and three flavors of light quarks to perform the
chiral and continuum extrapolations.  
Our prediction for the physical
value of the $I=2$ $\pi\pi$ scattering length is $m_\pi
a_{\pi\pi}^{I=2} = -0.04330 \pm 0.00042$, which agrees within uncertainties
with the (non-lattice) determination of CGL~\cite{Colangelo:2001df},
but  we emphasize
once again that our result rests on the assumption that the
fourth-root trick recovers the correct continuum limit of QCD.  In
Table~\ref{tab:vardet} and fig.~\ref{fig:barchart} we offer a
comparison of our prediction with other determinations.
\begin{table}[!ht]
\caption{\it A compilation of the 
various calculations and predictions for the $I=2$ $\pi\pi$ scattering
length.  The prediction made in this paper is labeled NPLQCD (2007).
Also included are the experimental value from
Ref.~\cite{Pislak:2003sv} (E 865 (2003)), the previous determination
by NPLQCD~\cite{Beane:2005rj} (NPLQCD (2005)), two indirect lattice
results from MILC~\cite{Aubin:2004fs,Bernard:2006wx} (the stars on the
MILC results indicate that these are not lattice calculations of the
$I=2$ $\pi\pi$ scattering length but rather a hybrid prediction which
uses MILC's determination of various low-energy constants together
with the Roy equations), and the Roy equation determination of Ref.~\cite{Colangelo:2001df}
(CGL (2001)).}
\label{tab:vardet}
\begin{ruledtabular}
\begin{tabular}{cc}
{} &  $m_\pi\ a_{\pi\pi}^{I=2}$ \\
\hline
$\chi$PT (Tree Level) & $-0.04438$ \\
NPLQCD (2007) & $-0.04330 \pm 0.00042$ \\
E 865 (2003) & $-0.0454\pm0.0031\pm0.0010\pm0.0008$ \\
NPLQCD (2005) & $-0.0426\pm0.0006\pm0.0003\pm0.0018$ \\
MILC (2006)* & $-0.0432\pm0.0006$ \\
MILC (2004)* & $-0.0433\pm0.0009$ \\
CGL (2001) & $-0.0444 \pm 0.0010$ \\
\end{tabular}
\end{ruledtabular}
\end{table}
What has enabled such an improvement in precision over our previous
result on the coarse MILC lattices is the recent understanding of the
lattice-spacing artifacts accomplished with mixed-action chiral
perturbation theory. 

While it will be quite useful to have results at another lattice
spacing and at another lattice volume, we have reached the level of
precision where we require knowledge of isospin violating effects in
order to further reduce the uncertainty in the physical $\pi\pi$
scattering lengths; i.e. those that can be compared to experiment.
One somewhat surprising result of our analysis is that one of the
dominant sources of systematic uncertainty in our calculation is due
to residual chiral symmetry breaking in the domain-wall valence quarks
for the lattice parameters we have chosen.  Clearly this systematic
can be reduced by improving our choice of domain-wall parameters.

Lattice QCD is currently in a precision age insofar as single-particle
properties are concerned. The precise prediction for the intrinsic
two-particle property presented here is a remarkable demonstration of
the power of combining a lattice QCD calculation with the
model-independent constraints of chiral perturbation theory.

%%%%%%%%%%%%%%%%%%%%%%%%%%%%%%%%%%%
\section{Acknowledgments}

\noindent We thank R.~Edwards for help with the QDP++/Chroma
programming environment~\cite{Edwards:2004sx} with which the
calculations discussed here were performed. AWL would like to thank
Claude Bernard for providing a program to determine meson mass
splittings in lattice units.  The computations for this work were
performed at Jefferson Lab, Fermilab, Lawrence Livermore National
Laboratory, National Center for Supercomputing Applications, and
Centro Nacional de Supercomputaci\'on (Barcelona, Spain).  We are
indebted to the MILC and the LHP collaborations for use of their
configurations and propagators, respectively.  The work of MJS was
supported in part by the U.S.~Dept.~of Energy under Grant
No.~DE-FG03-97ER4014. The work of KO was supported in part by the
U.S.~Dept.~of Energy contract No.~DE-AC05-06OR23177 (JSA) and contract
No.~DE-AC05-84150 (SURA) as well as by the Jeffress Memorial Trust,
grant J-813.  The work of AWL was supported in part by the
U.S.~Dept.~of Energy grant No.~DE-FG02-93ER-40762.  The work of SRB
and AT was supported in part by the National Science Foundation under
grant No.~PHY-0400231.  Part of this work was performed under the
auspices of the US DOE by the University of California, Lawrence
Livermore National Laboratory under Contract No. W-7405-Eng-48. 
The work of AP was partly supported by the EU contract
FLAVIAnet MRTN-CT-2006-035482, by the contract FIS2005-03142 from MEC 
(Spain) and FEDER and by the Generalitat de
Catalunya contract 2005SGR-00343.

%\vfill\eject

\end{document}